# Towards Guidelines for Assessing Qualities of Machine Learning Systems

Julien Siebert[1], Lisa Joeckel[1], Jens Heidrich[1], Koji Nakamichi[2],
Kyoko Ohashi[2], Isao Namba[2], Rieko Yamamoto[2], and Mikio Aoyama[3]

[1] Fraunhofer IESE, Kaiserslautern, Germany
{julien.siebert, lisa.joeckel, jens.heidrich}@iese.fraun-
hofer.de
[2] Fujitsu Laboratories Ltd., Kawasaki, Japan
{nakamichi, ohashi.kyoko, namba, r.yamamoto}@fujitsu.com
[3] Nanzan University, Nagoya, Japan
mikio.aoyama@nifty.com

**Abstract.** Nowadays, systems containing components based on machine learning (ML) methods are becoming more widespread. In order to ensure the intended behavior of a software system, there are standards that define necessary quality aspects of the system and its components (such as ISO/IEC 25010). Due to the different nature of ML, we have to adjust quality aspects or add additional ones (such as trustworthiness) and be very precise about which aspect is really relevant for which object of interest (such as completeness of training data), and how to objectively assess adherence to quality requirements. In this article, we present the construction of a quality model (i.e., evaluation objects, quality aspects, and metrics) for an ML system based on an industrial use case. This quality model enables practitioners to specify and assess quality requirements for such kinds of ML systems objectively. In the future, we want to learn how the term quality differs between different types of ML systems and come up with general guidelines for specifying and assessing qualities of ML systems.

**Keywords:** Machine Learning, Software Quality, Quality Evaluation.

## 1 Introduction

The digital transformation enables digital products and services that are based on data or on models derived from data. The construction of these models for algorithmic decision-making is increasingly based on artificial intelligence (AI) methods, which enable innovative solutions such as automated driving or predictive maintenance. Our research focuses on ML systems, i.e., software-intensive systems containing one or more components that use models built with ML methods. The functionality of these components is not defined by the programmer in the classical way, but is learned from data. Developing and operating ML systems raises new challenges in comparison to "classical" software engineering [1, 2]. First, the behavior is fundamentally different from traditional software: The relationship between the input and the outcome of the model is only defined for a subset of the data, which leads to uncertainty in model outcomes for previously unseen data. Second, common development principles from software engineering, such as encapsulation and modularity, have to be rethought, e.g., neural networks cannot simply be cut into smaller sub-nets and reused as modules. Third, the development and integration of ML components is a multi-disciplinary approach: It requires knowledge about the application domain, knowledge about how to construct ML models, and finally, knowledge about software engineering. Fourth, quality assurance, and specifically testing, works differently than in traditional software. This is because ML targets problems where the expected solution is inherently difficult to formalize [3].

In order to ensure the intended quality of a software system, there are standards that define necessary quality aspects of the system and its components. For instance, ISO/IEC 25010 [4] defines quality models for software and systems; i.e., a hierarchy of quality aspects of interest and how to quantify and assess them. Due to the different nature of ML, these quality models cannot be applied directly as they are. Some have to be adjusted in their definition (e.g., reusability of ML systems) and some need to be added (e.g., trustworthiness). We also have to be very precise about which quality aspect is relevant for which part of the overall system. For instance, in an ML system, the algorithms executing the model play a far less significant role than the data used for training and testing. For developing meaningful quality models, it is necessary to understand the application context of the use case and what kind of ML method is used.

In this article, we present the construction of a concrete quality model for an ML system based on an industrial use case. In this paper, we will first discuss related work and summarize the gaps that we would like to close with our contribution. Second, we will define the different views one can take on an ML system and relevant measurement objects, which will have to be evaluated for a specific use case and application context. Third, we will describe our general methodological approach for quality modeling of ML systems based on an industrial use case. This includes specifically the quality model containing all relevant quality aspects and concrete metrics for each measurement object of interest. This quality model enables practitioners to specify and assess quality requirements for such kinds of ML systems. Fourth, we will discuss the usefulness of the identified quality aspects based on an evaluation together with experts from industry. Last, we will present major lessons learned and give an outlook on future research, where we want to find out how the term quality differs between different types of ML systems (e.g., based on the ML method used or the way the ML component is integrated into the overall system). This will helps us come up with more general guidelines for specifying and assessing qualities of ML systems.

## 2    Related Work

To build a quality model, it is first necessary to define quality attributes. In the literature, some quite generic quality models for software and systems, such as ISO/IEC 25010 [4] or ISO/IEC 8000 [5], can be found. These standards propose different quality attribute definitions grouped into several categories with a decomposition structure (e.g., Product quality is decomposed into eight attributes, such as Functional Suitability, which is decomposed into sub-attributes, such as Functional Correctness). With the advance and the widespread adoption of ML methods, new and more specific quality proposals have emerged (such as the EU Ethics Guidelines for Trustworthy AI [6], the German DIN SPEC 92001 [7], or the Japanese QA4AI consortium [8]) as well as certification guidelines [9, 10].

Some of the new quality attributes are rather generic, so they cover not only ML but also other AI disciplines. These include:

- **Transparency and accountability** (e.g., reproducibility, interpretability and explainability, auditability, minimization, and reporting of negative impact)
- **Diversity, non-discrimination, fairness**, as well as **societal and environmental well-being** (e.g., avoidance of unfair bias, accessibility and universal design, stakeholder participation, sustainability, and environmental friendliness)
- **Security, safety, data protection** (e.g., respect for privacy, quality and integrity of data, access to data, and ability to cope with erroneous, noisy, unknown, and adversarial input data)
- **Technical robustness, reliability, dependability** (e.g., correctness of output, estimation of model uncertainty, robustness against harmful inputs, errors, or unexpected situations)
- **Human agency and oversight, legal and ethical aspects** (e.g., possibility of human agency and human oversight, respect for fundamental rights)

Some quality attributes are more specific to interactive and embodied AI (like assistants or robots), such as intelligent behavior and personality [8, 11, 12]. The quality attributes are applied to measurement objects. These objects can represent processes, products, impacted users, or external objects.

It is not uncommon to describe the system under study in terms of different views and measurement objects and to group the different quality attributes under these views/objects. For example, in [3], the authors define a set of quality attributes, such as correctness (i.e., goodness of fit), robustness, efficiency, etc. They also relate these quality attributes to different views/objects: data, learning program, and framework (e.g., Weka, TensorFlow). In [13], the authors distinguish between three main quality aspects, namely service quality, product quality, and platform quality. They also describe different views/objects on the system: the training dataset, the neural network, the hyper parameters, "the inference in vivo" (corresponding to the decision outputted by the ML component at runtime) and the machine learning platform. DIN-SPEC 92001 also provides a description in terms of views/objects: data, model, platform, and environment. As a last example, the authors in [8] provide five main quality aspects related to views/objects, namely data integrity, model robustness, system quality, process agility, and customer expectation, including a total of 49 quality sub-attributes.

In the literature, we see that a consensus exists around what quality aspects need to be measured. However, the naming of the quality attributes and the naming of the measurement objects (or the views) has not yet stabilized.

The same conclusion can be reached for quality attributes related to the process. Process models related to data analysis methods (such as knowledge engineering, data mining, ML, etc.) have been around for decades [14, 15]. In the last years, more case studies and literature reviews have been conducted to assess the challenges perceived by developers of ML components, as well as their processes and best practices [1, 3, 16, 17]. We see that there is a consensus on the definition of tasks, roles, and how the process should be organized for developing and operating ML components. However, it is less clear how the ML process impacts quality. Implementing quality improvement actions requires a good understanding of the process: which steps are performed, which people/roles are involved, which measurement objects are affected, etc.

We also see that, because the field of ML is large, the importance of certain quality attributes and metrics for quantifying them depend on the concrete context and use case and have to be addressed in different tasks of the process model used. For instance, the availability of a ground truth is one important factor (see **Fig. 1**): (a) If the full ground truth exists (as in the case of reinforcement learning, for example), then test oracles exist. Consequently, the quality mainly depends on the test oracle itself, and the quality can be safely measured using the available ground truth. (b) If only a partial ground truth exists (as in the case of semi-supervised or supervised learning), data quality and its representativeness have to be analyzed carefully. (c) If no ground truth exists (as in the case of unsupervised learning), the assumptions made by the learning algorithms and those made during the model evaluation play a significant role with respect to quality. The type of ML tasks that is performed (such as regression, classification, clustering, outlier detection, dimensionality reduction, etc.) also has an impact on the quality assessment. Each type of task is accompanied by corresponding quality metrics. For example, for classification tasks, the goodness of fit can be measured by accuracy, precision, recall, f-score, etc. [18], but for clustering, other measures are needed [19]. The metrics chosen will depend on the use case. For example, in the case of binary classification tasks, the cost of a false-positive may not be the same as the cost of a false-negative. Some metrics might not be compatible with one another, as is the case, for example, for fairness measures [20, 21].

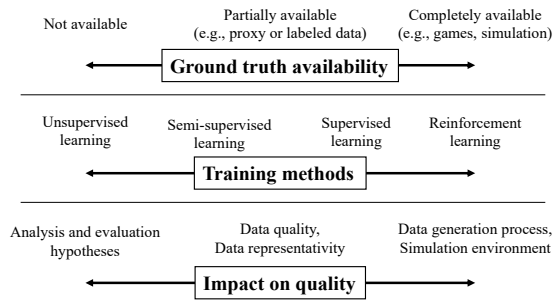

**Fig. 1.** The availability of ground truth data (labels) has a direct impact on the analysis or training methods used as well as on the definition of quality metrics and their assessment.

The literature provides a solid basis of relevant quality attributes, measurement objects, process models, etc. However, we see different gaps that have not been addressed so far: (1) There is a lack of unique and clear definitions of views on ML systems (e.g., what is the definition of a platform view in [7], or should hyper-parameters be included as a separate view). (2) Existing quality models are often too abstract to be of value for practitioners (e.g., in terms of proposed metrics) and require guidelines for tailoring to be applicable [22]. (3) The combination of and the relationship between quality attributes and related metrics have not been sufficiently investigated yet, and it is not clear whether they can be satisfied altogether. (4) Comprehensive development guidelines for quality-aware ML systems, which would bring together the different ML quality models, processes, and views, are largely missing or not made explicit.

In the remainder of this article, we will contribute mainly to closing the first two gaps. However, our overall research goal is aimed at coming up with comprehensive development guidelines for quality-aware ML systems.

## 3 Views on ML Systems

Many factors can influence the quality of a software system (code, hardware, development process, usage scenarios, etc.). In our approach, we tried to systematically identify groups of factors belonging together. We propose different "views" that would be helpful in categorizing quality attributes and corresponding quality metrics together with the objects to be measured. These views are: model view, data view, system view, infrastructure view, and environment view (see **Fig. 2** for an illustrative overview). Note that a given quality model may or may not use all the views, as the relevant ones are selected according to the use case.

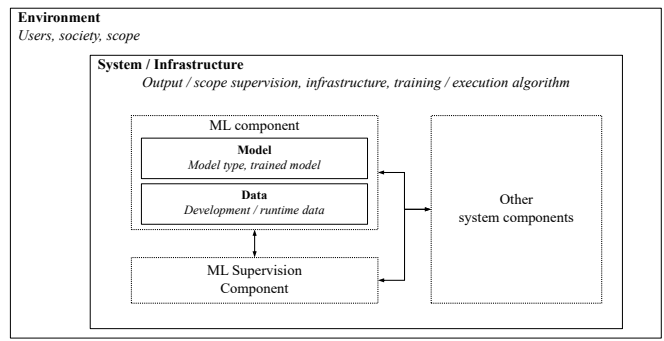

**Fig. 2.** Overview of the different views on the software system and measurement objects that influence the system's quality.

**Model View.** The model view is concerned with quality aspects belonging to what is called *a model* in machine learning. The model is the part that is trained on data in order to perform a given task (e.g., a classification, regression, dimensionality reduction, etc.). Note that an ML component normally does not contain only a single model but may be composed of several models, usually organized in a directed acyclic graph (also called a pipeline) [17]. The specificity of an ML component is the way it is built. We have to distinguish between the development phase (where the training and the evaluation of the pipeline is done) and the operation phase (where the artifacts created in the previous phase are deployed and used in production, i.e., at runtime), because these two phases may be implemented with different technologies (e.g., R/Python for learning, Web/Java on the application side) or have different quality demands (e.g., using a large quantity of data, operating under short latency, etc.).

In the model view, we have made a distinction between what we call a *model type* (e.g., decision trees, neural networks, etc.) and a *trained model* (e.g., a specific instance of a neural network trained on a specific dataset using a specific *training algorithm*). Again, the goal of this distinction is to separate quality aspects related to a specific object instance from those related to the object type. For example, the *appropriateness* of a given model applies to a *model type* (like the family of decision trees), whereas the *goodness of fit* applies to a specific trained instance. Note that we also separate the model from its *training algorithm* (i.e., the algorithm that takes *training data* and a *model type* as input and outputs a *trained model*) and its *execution algorithm* (i.e., the algorithm that takes *runtime data* and a *trained model* as input and outputs a decision, for instance a classification of inputted *runtime data*). The argumentation is that the training and execution algorithms are pieces of "classical" software whose quality aspects can be described and measured using existing standards.

**Data View.** The data view is concerned with the quality aspects related to the data. The term data here describes the data that is used as input for a model. We further distinguish between the *development data*, i.e., data used during the development phase to train the ML component, and *runtime data*, i.e., the dataset used during the operation phase. The distinction is made because these can be different physical objects, stored in different databases, potentially preprocessed or accessed differently during the development and operation phases. Therefore, different quality aspects apply either to each dataset separately or to both (for example, by comparing the representativeness of the *development data* with regard to the *runtime data*). We pushed the distinction even further concerning the *development data*. Indeed, the process of training an ML component requires splitting the *development data* into different subsets: the so-called training, validation, and test subsets. The *training subset* is used to determine the model parameters during training. The *validation subset* is used for hyper-parameter tuning (e.g., maximum depth of a decision tree). Finally, to provide an unbiased evaluation of the *trained model*, a *test subset* is used. Note that the *test subset* is supposed to be independent of the training and validation subsets. The way the *training*, *validation*, and *test subsets* are chosen have an impact on the quality of the evaluation of the trained model.

**System View.** First, an ML component is usually organized in a pipeline of tasks. Developing such a pipeline is by nature experimental. A given pipeline may be trained several times with different model types, training algorithms, or datasets. The way these sub-components are connected have an impact on quality (see, for example, the problem of data leakage [23]). Second, the ML component is part of a larger system, i.e., it consumes data from one or several sources and interacts with other ML or non-ML components. Since, a decision outputted by an ML component is always subject to uncertainty, and since wrong decisions might impact the system's overall quality, considering the flow of information from the system input through all components to the system output is important in order to understand the impact of the ML components' quality on the overall system behavior. Typical quality aspects related to the system view include, among others, data dependencies and feedback loops [2]. In our use case, the output of the ML component is monitored in order to detect and correct wrong decisions. This monitoring also has its own quality aspects that may be relevant for the use case at stake (e.g., monitoring effectiveness and efficiency).

**Infrastructure View.** What we call the infrastructure view is closely related to the system view. However, the view is here more focused on the quality aspects related to how the system is concretely implemented (e.g., hardware, training libraries). We decided to separate both views in order to highlight some specificities of ML components. For example, the efficiency of the *training* and *execution algorithms* is a quality attribute that belongs to this view. The same applies to the suitability of the infrastructure either for training or for executing ML components. For example, current trained deep learning models used for natural language processing are several gigabytes in size, and require several days (or weeks) of training on dedicated hardware machines (GPU clusters). The trained model cannot be executed on embedded devices due to computational and storage limitations.

**Environment View.** The *environment* consists of elements that (1) are external to the system under consideration, and (2) interact either directly or indirectly with the system. This includes the users. For ML systems, several environmental aspects may have a direct influence on the quality. These are, for example, aspects causing quality deficits in the data. This is strongly related to the notion of concept drift. Since an ML component is built for and tested in a given context of use (or target application scope), its quality will decrease when this context changes [24]. A self-adapting ML component dependent on the environment also raises further quality-related challenges (see, for example, the problems faced by the Microsoft chatbot Tay). Viceversa, an ML component can also have an impact on its environment, e.g., in terms of resource usage or societal discrimination [6].

## 4    Quality Modeling for ML Systems

Many aspects can influence the quality of a software component using ML and deriving a quality model for a specific use case may not be trivial. In this section, we will illustrate the approach we used to derive a quality model. The use case and the resulting quality model will be presented as well.

Our approach can be summarized as follows. We started by defining the relevant use case. This is usually done through interviews with the appropriate stakeholders. During these interviews, we made it clear what type of problem the ML component was supposed to solve (e.g., classification), what the intended application scope was (i.e., in which context the ML component should be used, what could change and how often), and whether some ground truth is available. We then used the views defined in the previous section to select pertinent measurement objects for the use case. From that point on, we selected quality attributes of interest and derived corresponding metrics (see **Fig. 3**).

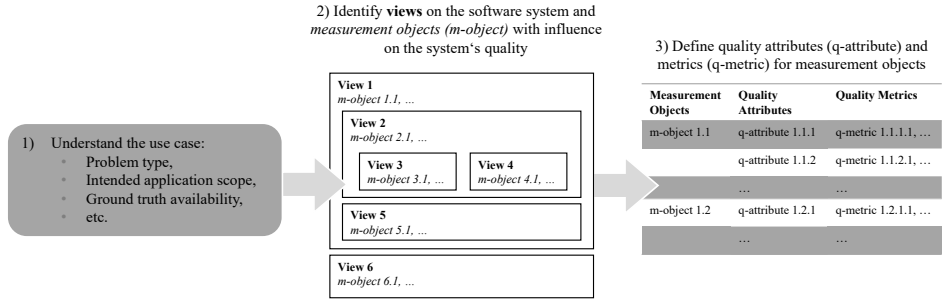

**Fig. 3.** Overview of the approach used to derive a specific quality model.

The industrial use case was as follows. Fujitsu's Accounting Center receives purchase order requests (POR) in digital form that need to be categorized for further treatment. This task was traditionally done by human operators and is now performed by an ML component. This component is trained with labeled examples of POR that used to be categorized in the past by human operators. The goal of the ML system is to reduce operating cost at an acceptable level of classification accuracy (comparable to humans). During the interviews, some quality issues were mentioned. First, the development and operation of ML components are complex and associated with high cost and risks. Several areas of expertise come into play when developing and operating an ML system. This may lead to potential communication and coordination problems. Furthermore, if wrong classifications do occur, finding the root cause of such a failure is not trivial. In order to deal with this, a monitoring system was implemented in conjunction with a correction engine based on expert rules. The ML component was re-trained when too many categorization failures were detected. This use case illustrates some of the quality issues encountered while developing and operating software systems using ML components.

**Table 1** presents the quality model we derived for the use case described above. For each view on the ML system, we defined a set of relevant quality attributes and corresponding measurement objects. For each attribute, we either give examples of concrete metrics for objectively evaluating the quality, or, if this is not possible, define examples for items one would have to check in order to address the respective quality. The quality model was designed to be specific enough to address the described use case appropriately (including supervision- and classification-related quality attributes), but also contains generic elements to allow it to be applied to other (similar) use cases (such as most attributes related to data and model).

How these measures could be aggregated in order to get an overall evaluation of the ML system and how to define quality improvement strategies for the use case are issues beyond the scope of this article.

**Table 1.** Overview of the derived quality model

| View | Measurement Object | Quality Attribute | Example Quality Metrics and Checklists |
|---|---|---|---|
| Model | Model type | **Appropriateness:** Degree to which the *model type* is appropriate for the current task (e.g., classification, etc.) and can deal with the current data type (e.g., numerical, categorical). | Prerequisites for model type. |
| | Trained model | **Development correctness (Goodness of Fit):** Ability of the model to perform the current task measured on the development dataset. | Precision, Recall, F-score, etc. for training. |
| | | **Runtime correctness (Goodness of Fit):** Same as above measured on the runtime dataset. | Precision, Recall, F-score, etc. at runtime. |
| | | **Relevance (Bias-Variance tradeoff):** Degree to which the model achieves a good bias-variance trade-off (neither underfitting nor overfitting the data). | Variance of cross-validation goodness of fit. |
| | | **Robustness:** Ability of the model to handle noise or data with missing values and still make correct predictions. | Equalized Loss of Accuracy (ELA). |
| | | **Stability:** Degree to which a trained model generates repeatable results when it is trained on different subsets of the training dataset. | Leave-one-out cross-validation stability. |
| | | **Fairness:** Ability of the model to output fair decisions. | Equalized odds. |
| | | **Interpretability:** Degree to which the trained model can be interpreted by humans. | Complexity metrics (e.g., no. of parameters, depth) |
| | | **Resource utilization:** Resources used by the model when it is already trained. | Required storage space. |
| Data | Development data | **Representativeness:** Degree to which the data is representative of the statistical population. | Statistical tests (e.g., two-sample t-test, etc.). |
| | | **Correctness:** Degree to which the data is free from errors. | Outlier detection metrics (e.g., Z-score). |
| | | **Completeness:** Degree to which the data is free from missing values. | No. of missing values. |
| | | **Currentness:** Degree to which the data is up to date w.r.t. the current task. | Age of data. |
| | | **Intra-Consistency:** Consistency of the data within a dataset, e.g., the data does not contradict itself or the formatting is consistent. | Value ranges, word counts. |
| | | **Train/Test Independence:** Degree to which the *training* and *test subsets* are independent of one another. | Statistical tests (e.g., two-sample t-test, etc.). |
| | | **Balancedness:** Degree to which all classes (labels) are equally represented in the dataset. | Ratio of classes. |
| | | **Absence of Bias:** Degree to which the data is free from bias against a given group. | Ratios of groups. |
| | Development and runtime data | **Inter-Consistency:** Consistency between different datasets, e.g., formatting, sampling methods used. | Value ranges, crosswise outlier detection metrics. |
| Environment | Training process | **Environmental Impact:** Degree to which the training process impacts the environment. | Energy consumption. |
| | Society | **Social Impact:** Degree to which the ML component impacts society. | Impact on employees. |
| | Scope | **Scope Compliance:** Degree to which the application of the ML component respects its intended scope of use. | Value ranges, novelty detection metrics. |
| System | Output supervision | **Effectiveness:** Degree to which the output supervision algorithm detects false outcomes of the ML component | False positive/negative detection rate. |
| | | **Supervision Overhead / Efficiency:** Resources used for monitoring a given ML component | Time, memory used, etc. |
| | Scope supervision | **Effectiveness:** Degree to which the scope supervision algorithm detects context changes. | No. of out-of-scope cases. |
| | | **Supervision Overhead / Efficiency:** Resources used for monitoring the application scope. | Time memory used, etc. |
| | Other non-ML components | Here we refer to the relevant subset of the quality attributes of the standard ISO/IEC 25010, which are not listed here for space reasons. | |
| Infrastructure | Infrastructure | **Infrastructure Suitability:** Degree to which the infrastructure matches the ML component needs (e.g., in terms of hardware type, computation capability, bandwidth, memory, etc.) | Computational and storage capabilities. |
| | Training algorithm | **Training Efficiency:** Resources used for training a given model. | Time, memory used, etc. |
| | Execution algorithm | **Execution Efficiency:** Resources used for executing a given trained model. | Time, memory used, etc. |

# 5    Discussion

In this article, we first proposed a categorization of quality attributes as well as measurement objects in the form of different views/objects. This classification is the result of a literature-based review, discussions with industrial partners, and our own experience in ML component development. To scientifically assess and consolidate a useful and systematic grouping of quality attributes for ML systems (as well as measurement objects), several iterations will be necessary (e.g., case study, systematic literature review, mapping study).

We also derived a quality model specifically tailored for a given use case. The definition and relevance of the quality attributes were first discussed internally in a workshop with experts. Later, three case studies with a focus on requirements engineering for ML systems were conducted by Fujitsu Laboratories [25]. In this paper, the authors present the overall requirements engineering process, but do not go into the details of the quality model presented. The performed case studies confirmed that the quality attributes identified were valid and meaningful for ML developers, especially in the context of requirements specification.

In terms of limitations, we see three main aspects:

1.  We did not address process-related aspects yet, i.e., what qualities have to be assured in which activity and handled by which role. We believe that the proposed views/objects can help to establish a mapping between roles (e.g., Data Scientist, Data Engineer, etc.) and quality attributes or metrics. For example, Data Scientists are usually in charge of building models, and are in direct line when it comes to measuring the impact of data quality on the models' outcomes. However, Data Engineers are the ones that can usually implement new data quality improvement actions. Architects with a good understanding of ML will be needed in order to solve problems on the system level.

2.  The identified views may be incomplete and currently focus more on the later stages of CRISP-DM, missing the stages of Business Understanding (i.e., ML requirements) and Data Understanding, and their related measurement objects. For example, Data Understanding is by nature rather experimental and the artifacts produced at this stage usually consist of a set of decisions (e.g., which data preparation algorithm to choose) and may be accompanied by code snippets or visualizations (e.g., notebooks, reports). An open question is whether the views should be augmented with new measurement objects (such as specific ML requirements documents, experiment reports or notebooks, etc.) or whether another classification direction based upon processes is needed.

3.  Finally, our viewpoint for defining the quality model was more from the data science perspective. Integration with classical software/system engineering qualities (such as those defined by ISO/IEC 25010) is missing. There is as yet no consensus on the naming of ML-related quality attributes. Furthermore, whereas some of the proposed attributes can be easily classified under existing ISO/IEC 25010 ones (e.g., the model's Goodness of Fit could potentially belong to Functional Correctness), others may be more difficult to classify (such as Scope Compliance). Whether the ISO/IEC 25010 is the right framework for ML components is still an open issue.

# 6      Lessons Learned and Conclusions

This article presented how we constructed a concrete quality model for an ML system based on an industrial use case. We are completely aware that the model we developed is quite specific to the case and that other use cases may require different quality aspects and, in consequence, different metrics. However, we would like to share an excerpt from the lessons we learned from following the described methodological approach. Even though some of these are known from other fields, we nonetheless think it is worth mentioning them in the context of developing ML systems:

1. Context and use case must be clear. As pointed out before, there are many application fields and potential ML-based solutions available. It is very important to be as clear as possible about the general application context. ML models should never ever be used just for the sake of being fancy, but always because there is the profound assumption that they will add concrete value for the application context. The quality aspects that are important mainly depend on this.
2. Iterative approach: The ML model, its application context, and its use case have to be adjusted over time and some initial assumptions will turn out to be false. Therefore, it is important to follow an iterative approach when developing the ML system and to be able to quickly identify dead ends and take different paths. Having a clear picture of what quality aspects are important and how to quantify them is crucial for this, as it allows us to immediately see whether we can fulfill them with our solution path.
3. Multidisciplinary work: As we stated at the beginning of this article, different kinds of knowledge must come together for developing quality-aware ML systems. For instance, a data scientist knows how to measure the fairness or stability of the trained model, a software/system engineer knows how to assure the quality of the overall system, and a domain expert knows whether the ML system really solves the problem better than a traditional software system.
4. The devil is in the details: We learned that it is easy to talk about abstract generic quality aspects, such as those defined by ISO/IEC 25010, on a high level. To define meaningful quality aspects, we had to break them down into concrete qualities of measurement objects and define how to operationalize these aspects with metrics.
5. Quality-aware process/guidelines: Even though there are defined processes for ML model building (such as CRISP-DM) and for software engineering (such as rich and agile processes) with elaborate practices for improvement (such as DevOps approaches), an integrated process is missing, nor do guidelines exist on how to bring everything together with a clear focus on the quality of ML systems.

Regarding future work, we first plan to perform more case studies to empirically validate the different quality aspects in more detail, specifically their relevance for practitioners and how to deal with them in different process stages. Second, we want to apply this method to other ML problems (like regression, or unsupervised problems) and learn about the impact on the quality model. Third, we intend to package our insights into development guidelines for quality-aware ML systems.